\documentclass[12pt]{article}
\usepackage[utf8]{inputenc}
\usepackage{amsmath,amssymb}
\usepackage{hyperref}
\usepackage{cite}
\usepackage{braket} 
\usepackage[a4paper,margin=2.5cm]{geometry}
\title{On the Quantization of the Electromagnetic Field with Magnetic Monopoles}
\author{Kanan Anwar \\[1ex]
\small École Polytechnique de Bruxelles, ULB}
\date{September 2025}

\begin{document}
\maketitle

\begin{abstract}
We present a covariant framework for the quantization of the electromagnetic field in the presence of magnetic monopoles. Building on the two-potential formalism of Cabibbo and Ferrari, which treats electric and magnetic sources on equal footing and reveals a $U(1) \times U(1)$ gauge symmetry, we extend the theory into the quantum domain. Using the Gupta–Bleuler procedure in the Feynman-'t Hooft gauge, we construct the physical Hilbert space and eliminate negative-norm states. The resulting theory predicts, in addition to conventional photons, the existence of dual photons associated with magnetic charges. We discuss the role of these dual excitations and their possible relevance in the broader context of electromagnetic duality and gauge theories.
\end{abstract}

\section{Introduction}

The idea of magnetic monopoles has long occupied a central place in theoretical physics. In 1931, Dirac showed that their existence would not only be consistent with Maxwell’s equations but would also provide an elegant explanation for the quantization of electric charge, through the famous condition \textit{eg = n$\hbar$}. Despite this remarkable theoretical insight, monopoles have never been observed experimentally, and the standard formulation of quantum electrodynamics (QED) does not accommodate them in a manifestly covariant way. A more symmetric approach was proposed by Cabibbo and Ferrari\cite{cabibbo} in 1962, who introduced a two-potential formalism with four-vectors $(A_\mu, T_\mu)$ representing the electric and magnetic sectors, respectively. This construction eliminates the need for the Dirac string, restores Lorentz covariance, and makes the electromagnetic duality explicit through a $U(1) \times U(1)$ gauge invariance. The formalism has been revisited in various contexts, including generalized electromagnetism, effective models of monopoles, and duality-invariant field theories. While the classical properties of this framework are well established, its systematic quantization has received comparatively less attention. In particular, the structure of the Hilbert space and the treatment of unphysical polarization states are nontrivial in the presence of two independent gauge fields. The Gupta–Bleuler\cite{gupta} procedure, originally developed for conventional QED, provides a consistent method for quantizing gauge theories while eliminating non-physical states. Applying this method to the Cabibbo–Ferrari formalism, we find that the resulting quantum theory naturally contains, besides the familiar photon, a new gauge boson namely the dual photon. The aim of this work is not to propose a fully realistic extension of QED, but rather to present a covariant construction of electromagnetic field quantization in the presence of monopoles. Our analysis highlights how the dual photon emerges from the doubled Gupta–Bleuler structure and discusses its possible interpretation in the context of modern gauge theories. 
\section{Magnetic monopoles}

The Maxwell equations in the presence of both electric and magnetic charges take the symmetric form
\begin{align}
    \nabla \cdot \mathbf{E} &= \rho_e, \\
    \nabla \cdot \mathbf{B} &= \rho_m, \\
    \nabla \wedge \mathbf{E} &= -\frac{\partial \mathbf{B}}{\partial t} - \mathbf{J}_m, \\
    \nabla \wedge \mathbf{B} &= \mathbf{J}_e +  \frac{\partial \mathbf{E}}{\partial t},
\end{align}
and with the generalized scalar and vector potentials, the fields $\mathbf{E}$ and $\mathbf{B}$ can be expressed as
\begin{equation}
\mathbf{B} = -\nabla \Phi - \frac{\partial \mathbf{T}}{\partial t} + \nabla \wedge\mathbf{A}, \quad
\mathbf{E} = -\nabla V - \frac{\partial \mathbf{A}}{\partial t} - \nabla \wedge \mathbf{T}
\end{equation}
with $\Phi$ being the magnetic potential and $V$ the electric potential. By definition, sources are encoded in the four-currents
\begin{equation}
J_m^\mu = \left( \rho_m, \mathbf{J}_m \right)
\end{equation}
\noindent and
\begin{equation}
J_e^\mu = \left( \rho_e, \mathbf{J}_e \right).
\end{equation}  
which satisfy the equations below
\begin{equation}
\partial_\mu J_m^\mu = 0, \quad
\partial_\mu J_e^\mu = 0
\end{equation}
We introduce two four-potentials as
\begin{equation}
A^\mu = \left( V, \mathbf{A} \right), \quad
T^\mu = \left( \Phi, \mathbf{T} \right),
\end{equation}
such that the following relations are verified
\begin{equation}
\Box A^\mu = J_e^\mu, \quad
\Box T^\mu = J_m^\mu,
\end{equation}
where \(\Box\) denotes the d'Alembertian operator. The fields admit the following gauge transformations:
\begin{align}
    A^\mu &\;\to\; A^\mu + \partial_\mu \Lambda_1, \quad
    T^\mu \;\to\; T^\mu + \partial_\mu \Lambda_2 .
\end{align}
These independent gauge transformations reveal a U(1) × U(1) symmetry structure, reflecting the electromagnetic duality between electric and magnetic charges. By imposing the Lorenz gauge condition,
\begin{equation}
    \partial_\mu A^\mu = 0, \quad
    \partial_\mu T^\mu = 0 ,
\end{equation}
one finds
\begin{align}
    \Box \Lambda_1 = 0, \quad
    \Box \Lambda_2 = 0,
\end{align}
so that the solutions to equations (10) are uniquely determined. From that we can define the Lorentz invariant Lagrangian in terms of the following gauge invariant field tensors
\begin{equation}
F^{\mu \nu} = \partial^\mu A^\nu - \partial^\nu A^\mu, \quad
K^{\mu \nu} = \partial^\mu  T^\nu - \partial^\nu T^\mu.
\end{equation}
This construction differs from the earlier approach of Shanmugadhasan \cite{shanmugadhasan}, where a single generalized tensor was defined by combining both potentials:
\begin{align}
    F^{\mu \nu} = \partial^\mu A^\nu - \partial^\nu A^\mu - \varepsilon^{\mu \nu \alpha \beta}\partial_\alpha T_\beta.
\end{align}
By starting from the standard Lagrangian of classical electrodynamics, we have
\begin{equation}
\mathcal{L} = -\frac{1}{4} F^{\mu \nu} F_{\mu \nu} -\frac{1}{4} K^{\mu \nu} K_{\mu \nu}.
\end{equation}
and if we consider the interactions between the charges and the electromagnetic field,
\begin{equation}
\mathcal{L} = -\frac{1}{4} F^{\mu \nu} F_{\mu \nu} -\frac{1}{4} K^{\mu \nu} K_{\mu \nu} - A_{\nu} J^{\nu}_{e} - T_{\nu} J^{\nu}_{m}.
\end{equation}
Note that we are using the (-,+,+,+) convention for the metric tensor. We immediately see that the Lagrangian depends on 4 variables :
\begin{equation}
\mathcal{L} = \mathcal{L}(A_{\nu}, T_{\nu}, \partial_\mu A_{\nu}, \partial_\mu T_{\nu}).
\end{equation}
The associated action is defined as
\begin{align}
S = \int d^4x \, \mathcal{L}(A_{\nu}, T_{\nu}, \partial_\mu A_{\nu}, \partial_\mu T_{\nu})
\end{align}
by demanding that the action be stationary under variations of the fields, we have
\begin{equation}
\delta S = \int d^4x \,  \delta \mathcal{L}(A_{\nu}, T_{\nu}, \partial_\mu A_{\nu}, \partial_\mu T_{\nu}) = 0,
\end{equation}
we get the associated Euler-Lagrange equations. By expanding the variation of the Lagrangian, we get the following expression :
\begin{equation}
\delta \mathcal{L} = \frac{\partial \mathcal{L}}{\partial A_{\nu}} \delta A_{\nu} + \frac{\partial \mathcal{L}}{\partial T_{\nu}} \delta T_{\nu} + \frac{\partial \mathcal{L}}{\partial (\partial_\mu A_{\nu})} \delta (\partial_\mu A_{\nu}) + \frac{\partial \mathcal{L}}{\partial (\partial_\mu T_{\nu})} \delta (\partial_\mu T_{\nu}).
\end{equation}
Using integrating by parts, we easily get that 
\begin{equation}
\delta S = \int d^4x \left( \frac{\partial \mathcal{L}}{\partial A_{\nu}} \delta A_{\nu} + \frac{\partial \mathcal{L}}{\partial T_{\nu}} \delta T_{\nu} - \partial_\mu \left( \frac{\partial \mathcal{L}}{\partial (\partial_\mu A_{\nu})} \right) \delta A_{\nu} - \partial_\mu \left( \frac{\partial \mathcal{L}}{\partial (\partial_\mu T_{\nu})} \right) \delta T_{\nu} \right) = 0,
\end{equation}

\noindent by factorizing out \( \delta A_{\nu} \) and \( \delta T_{\nu} \), we easily obtain the desired Euler-Lagrange equations.

\begin{equation}
\frac{\partial \mathcal{L}}{\partial A_{\nu}} - \partial_\mu \left( \frac{\partial \mathcal{L}}{\partial (\partial_\mu A_{\nu})} \right) = 0, \forall \ \delta A_\nu
\end{equation}
\noindent and,
\begin{equation}
\frac{\partial \mathcal{L}}{\partial T_{\nu}} - \partial_\mu \left( \frac{\partial \mathcal{L}}{\partial (\partial_\mu T_{\nu})} \right) = 0, \forall\ \delta T_\nu,
\end{equation}
By computing the derivatives, we can get the modified Maxwell's equations in covariant form. The first term in each equation, representing the derivative with respect to the potential, simplifies to the 4-current, leading directly to the following expressions.

\begin{equation}
\frac{\partial \mathcal{L}}{\partial A_{\nu}} = -J^{\nu}_e, \quad
\frac{\partial \mathcal{L}}{\partial T_{\nu}} = -J^{\nu}_m.
\end{equation}
And for the second terms,
\begin{equation}
\frac{\partial \mathcal{L}}{\partial (\partial_\mu A_{\nu})} = -F^{\mu \nu},
\end{equation}
\noindent and similarly,
\begin{equation}
\frac{\partial \mathcal{L}}{\partial (\partial_\mu T_{\nu})} = -K^{\mu \nu}.
\end{equation}

\noindent Therefore, by substituting into equations (23) and (24), we obtain the complete set of Maxwell’s equations:

\begin{equation}
\partial_{\mu} F^{\mu \nu} = J^{\nu}_e, \quad
\partial_{\mu} \widetilde{F}^{\mu \nu} = 0,
\end{equation}
together with
\begin{equation}
\partial_{\mu} K^{\mu \nu} = J^{\nu}_m, \quad
\partial_{\mu} \widetilde{K}^{\mu \nu} = 0,
\end{equation}

\noindent where $\widetilde{F}^{\mu \nu}$ and $\widetilde{K}^{\mu \nu}$ denote the dual tensors of $F^{\mu\nu}$ and $K^{\mu\nu}$, respectively. These equations together provide the fully covariant formulation of Maxwell’s equations, including magnetic charges.
\section{Fields quantization}

In order to proceed with the canonical quantization of the theory, we first identify
the conjugate momenta associated with the two gauge fields. They are defined as
\begin{align}
\Pi^\mu_A = \dfrac{\partial\mathcal{L}}{\partial(\partial_0 A_\mu)}, \quad
\Pi^\mu_T = \dfrac{\partial\mathcal{L}}{\partial(\partial_0 T_\mu)}
\end{align}
being the canonical momenta associated with the fields $A^\mu$ and $T^\mu$, respectively.
We then find the Hamiltonian by doing a Legendre transformation :
\begin{equation}
\mathcal{H} = \Pi^\mu_A \dot A_\mu + \Pi^\mu_T \dot T_\mu - \mathcal{L}(A_\mu, T_\mu, \partial_v A_\mu, \partial_v T_\mu),
\end{equation}
where $\mathcal{L} = -\dfrac{1}{4} F^{\mu \nu} F_{\mu \nu} -\dfrac{1}{4} K^{\mu \nu} K_{\mu \nu}$ is the free Maxwell Lagrangian.
By calculating the canonical moments, we get that
\begin{equation}
    \Pi^i_A = \frac{\partial\mathcal{L}}{\partial(\partial_0 A_i)} = F^{i0}
\end{equation}
and for the temporal component 
\begin{equation}
    \Pi^0_A = \frac{\partial\mathcal{L}}{\partial(\partial_0 A_0)} = 0
\end{equation}
the same can be done for the $T^\mu$ field,
\begin{align}
    \Pi^i_T = \frac{\partial\mathcal{L}}{\partial(\partial_0 T_i)} = K^{i0}, \quad
    \Pi^0_T = \frac{\partial\mathcal{L}}{\partial(\partial_0 T_0)} = 0
\end{align}
Hence we cannot use this Lagrangian for the quantization of our fields, as the vanishing of the canonical momenta would prevent the construction of our Fock space.
We've imposed the Lorenz gauge at the beginning, but we will ignore that for now and use that condition later on. Let us elaborate a theory governed by the modified Lagrangian,
\begin{equation}
    \mathcal{L} = -\dfrac{1}{4} F^{\mu \nu} F_{\mu \nu} -\dfrac{1}{4} K^{\mu \nu} K_{\mu \nu} - \frac{1}{2 \lambda}(\partial_\mu A^\mu)^2 - \frac{1}{2 \lambda}(\partial_\mu T^\mu)^2
\end{equation}
which is slightly different from Maxwell's theory. Our Fock space associated to this Lagrangian contains non-physical polarization states, but we will eventually restrict it to the physical states only by using the Lorenz gauge. Therefore, we use a modified Lagrangian and by taking the Feynman-'t Hooft gauge ($\lambda = 1$), we can then apply the Gupta-Bleuler quantization
\begin{equation}
    \mathcal{L} = -\dfrac{1}{4} F^{\mu \nu} F_{\mu \nu} -\dfrac{1}{4} K^{\mu \nu} K_{\mu \nu} - \frac{1}{2}(\partial_\mu A^\mu)^2 - \frac{1}{2}(\partial_\mu T^\mu)^2
\end{equation}
It follows that the timelike component of the canonical momentum does not vanish and is given by
\begin{align}
    \Pi^0_A = -\partial_\mu A^\mu, \quad
    \Pi^0_T = -\partial_\mu T^\mu,
\end{align}
we therefore quantize the fields by imposing the commutations relations :
\begin{align}
    [\hat A_\mu(t,\mathbf{x}), \hat A_\nu(t,\mathbf{y})] &= [\hat \Pi^{\mu}_A(t, \mathbf{x}),\hat \Pi^{\nu}_A(t, \mathbf{y})] = 0 \\
    [\hat A_\mu(t, \mathbf{x}),\hat \Pi_{\nu, A}(t, \mathbf{y})] &= i \eta_{\mu \nu} \delta^{(3)}(\mathbf{x} - \mathbf{y}) \\
    [\hat T_\mu(t, \mathbf{x}),\hat \Pi_{\nu, T}(t, \mathbf{y})] &= i \eta_{\mu \nu} \delta^{(3)}(\mathbf{x} - \mathbf{y}) \\
    [\hat T_\mu(t,\mathbf{x}), \hat T_\nu(t,\mathbf{y})] &= [\hat \Pi^{\mu}_T(t, \mathbf{x}),\hat \Pi^{\nu}_T(t, \mathbf{y})] = 0
\end{align}
By expanding the potentials as a sum of plane waves in terms of creation and annihilation operators, we obtain:
\begin{align}
    A_\mu = \int \frac{d^3 k}{(2 \pi)^3} \frac{1}{\sqrt{2 \omega_k}} \sum_{\lambda = 0}^3\left(\epsilon_\mu^\lambda(\mathbf{k})\hat a_\lambda(\mathbf{k}) e^{-i\mathbf{k}.\mathbf{x}} + \epsilon_\mu^\lambda(\mathbf{k})\hat a_\lambda^\dagger(\mathbf{k})  e^{i\mathbf{k}.\mathbf{x}} \right) \\
    T_\mu = \int \frac{d^3 k}{(2 \pi)^3} \frac{1}{\sqrt{2 \omega_k}} \sum_{\lambda = 0}^3 \left(\epsilon_\mu^\lambda(\mathbf{k})\hat b_\lambda(\mathbf{k}) e^{-i\mathbf{k}.\mathbf{x}} + \epsilon_\mu^\lambda(\mathbf{k})\hat b_\lambda^\dagger(\mathbf{k})  e^{i\mathbf{k}.\mathbf{x}} \right)
\end{align}
with $\epsilon^\lambda_\mu$ being the polarization vector and the operators obeying
\begin{align}
    \hat a_\lambda(\mathbf{k}) \ket{0} = 0, \quad
    \hat b_\lambda(\mathbf{k}) \ket{0} = 0
\end{align}
$\ket{0}$ being the vacuum state. The operators satisfy the following commutation relations :
\begin{align}
    \left[\hat a_\lambda(\mathbf{k}), \hat a^\dagger_{\gamma}(\mathbf{q}) \right] = -\eta^{\lambda \gamma}(2\pi)^3 2\omega_k \delta^{(3)}(\mathbf{k - q}) \\
    \left[\hat b_\lambda(\mathbf{k}), \hat b^\dagger_{\gamma}(\mathbf{q}) \right] = -\eta^{\lambda \gamma}(2\pi)^3 2\omega_k \delta^{(3)}(\mathbf{k - q}),
\end{align}
However, these commutation relations lead to an issue: they give rise to negative norms. For instance, consider the timelike component ($\lambda = 0$) and the states
\begin{align}
    \ket{\psi} = \hat a^\dagger_0(\mathbf{k}) \ket{0}, \quad
    \ket{\phi} = \hat b^\dagger_0(\mathbf{k}) \ket{0}
\end{align}
taking the scalar product, we find
\begin{align}
    \braket{\psi | \psi} &= \bra{0} \hat a_0(\mathbf{k}) \hat a^\dagger_0(\mathbf{k}) \ket{0} \\
    &= \bra{0} \left[ \hat a_0(\mathbf{k}), \hat a^\dagger_0(\mathbf{k}) \right] \ket{0} \\
    &= -(2\pi)^3 2\omega_k \delta^{(3)}(\mathbf{0}) \braket{0 | 0} <0
\end{align}
and
\begin{align}
    \braket{\phi | \phi} &= \bra{0} \hat b_0(\mathbf{k}) \hat b^\dagger_0(\mathbf{k}) \ket{0} < 0
\end{align}
which is clearly unphysical. As mentioned before, our Fock space contains non-physical states, we must then construct a subspace of this Fock space corresponding to Maxwell's theory by using the Lorenz gauge
\begin{align}
    \partial_\mu A^\mu = 0, \quad
    \partial_\mu T^\mu = 0,
\end{align}
but this is incompatible with the commutation relations. Instead, we require that for any two physical states $\ket{\phi}$ and $\ket{\psi}$,
\begin{align}
    \bra{\phi} \partial_\mu A^\mu \ket{\psi} = 0, \quad
    \bra{\phi} \partial_\mu T^\mu \ket{\psi} = 0.
\end{align}
We then decompose the operator such that
\begin{align}
    \partial_\mu A^\mu = \partial_\mu A^\mu_+ + \partial_\mu A^\mu_-, \quad
    \partial_\mu T^\mu = \partial_\mu T^\mu_+ + \partial_\mu T^\mu_-
\end{align} 
to ensure that the vacuum is a physical state, we impose
\begin{align}
    \partial_\mu A^\mu_+ \ket{\psi} = 0
\end{align}
and
\begin{align}
    \partial_\mu T^\mu_+ \ket{\psi} = 0
\end{align}
with 
\begin{align}
    A^\mu_+ = \int \frac{d^3 k}{(2 \pi)^3} \frac{1}{\sqrt{2 \omega_k}} \sum_\lambda \epsilon^\mu_\lambda(\mathbf{k})\hat a_\lambda(\mathbf{k}) e^{-i\mathbf{k}.\mathbf{x}},\\
    T^\mu_+ = \int \frac{d^3 k}{(2 \pi)^3} \frac{1}{\sqrt{2 \omega_k}} \sum_\lambda \epsilon^\mu_\lambda(\mathbf{k})\hat b_\lambda(\mathbf{k}) e^{-i\mathbf{k}.\mathbf{x}},
\end{align}
which is precisely the Gupta-Bleuler condition. However, what we have obtained is essentially a doubled Gupta-Bleuler structure, revealing the presence of a new particle: the dual photon. The Hilbert space is therefore not yet fully restricted to physical states. Since we now have two distinct quantized fields rather than a single one, we distinguish between the physical photon states, denoted $\ket{\psi}_E$ and the dual photon states $\ket{\psi}_M$. We therefore expand these states in a Fock space basis
\begin{align} 
    \ket{\psi}_E = \ket{\psi_T}_E\ket{\phi}_E
\end{align}
with $\ket{\psi_T}_E$ containing transverse photons created by $\hat a_{1,2}^\dagger(\mathbf{k})$, while $\ket{\phi}_E$ contains longitudinal photons created by $\hat a_3^\dagger(\mathbf{k})$ and timelike photons which are created by $\hat a_0^\dagger(\mathbf{k})$. The same is done for the dual photons such that
\begin{align} 
    \ket{\psi}_M = \ket{\psi_T}_M\ket{\phi}_M,
\end{align}
and if we inject the expressions of the fields in the equations (56) and (57), we require that
\begin{align}
    \left(\hat a_3(\mathbf{k}) - \hat a_0(\mathbf{k}) \right)\ket{\phi}_E = 0
\end{align}
and
\begin{align}
    \left(\hat b_3(\mathbf{k}) - \hat b_0(\mathbf{k}) \right)\ket{\phi}_M = 0,
\end{align}
which means that timelike photons and longitudinal photons are not independent.
If we take the following scalar product, 
\begin{align}
    \bra{\phi}_E \left(\hat a_3(\mathbf{k}) - \hat a_0(\mathbf{k}) \right)^\dagger \left(\hat a_3(\mathbf{k}) - \hat a_0(\mathbf{k}) \right) \ket{\phi}_E &= \bra{\phi}_E \left(\hat a_3^\dagger(\mathbf{k})a_3(\mathbf{k}) - \hat a_0^\dagger(\mathbf{k}) \hat a_0(\mathbf{k}) \right) \ket{\phi}_E\\
    &= 0
\end{align}
meaning that the contributions from the timelike and longitudinal photons/dual photons cancel each other in our Hamiltonian. Therefore, the expectation value of the Hamiltonian, i.e. the energies carried by the photons and the dual photons in some state $\ket{\psi}$ is 
\begin{align}
    \bra{\psi} \hat H \ket{\psi} = \bra{\psi} \sum_{\lambda = 1}^2 \hat a^\dagger_\lambda(\mathbf{k}) \hat a_\lambda(\mathbf{k}) + \hat b^\dagger_\lambda(\mathbf{k}) \hat b_\lambda(\mathbf{k}) \ket{\psi}
\end{align}
 and as expected, the non-physical states are removed and inner products on our Hilbert space are positive definite. The most striking outcome of the quantization procedure is the emergence of the dual photon. This additional gauge boson arises naturally from the doubled Gupta–Bleuler structure: while conventional QED eliminates unphysical timelike and longitudinal modes, the two-potential formalism requires the same condition for both fields. As a result, two independent sets of transverse excitations survive, corresponding to photons and dual photons. The idea of two-photon theories has appeared several times in the literature. As early as 1966, Salam proposed a framework where the existence of magnetic monopoles could naturally be tied to the presence of a second photon, potentially linked to charge conjugation (C) violation \cite{Salam1966}. Additional support for this picture comes from models of confinement in compact QED and non-Abelian gauge theories, where effective dual photons play a role \cite{thooft}, as well as from dual-symmetric formulations of classical electromagnetism \cite{BliokhBekshaevNori}. Keller (2018) has also proposed a photon wave mechanics explicitly including monopoles, which requires distinct electric and magnetic propagating modes \cite{Keller2018}. Nevertheless, neither dual photons nor magnetic monopoles have been experimentally observed in spite of these theoretical advancements. However, some condensed matter systems, particularly spin ice materials, have been found to exhibit monopole-like excitations, where quasi particles effectively act as magnetic monopoles. Therefore, even though the dual photon is still only a theoretical prediction in the current framework, the fact that there are analogs of it in lab systems indicates that some aspects of monopole physics may already be relevant to experiments. In this way, the covariant quantization introduced here, in addition to offering a coherent mathematical framework, makes it possible to relate theoretical monopole dynamics to emergent phenomena in actual materials.


\end{document}